# Stratospheric Aerosol Particles Size Distribution
# Based on Multi-Color Polarization Measurements of the Twilight Sky

Oleg S. Ugolnikov[1] and Igor A. Maslov[1,2]
[1]Space Research Institute, Russian Academy of Sciences
Profsoyuznaya st., 84/32, Moscow 117997 Russia
[2]Moscow State University, Sternberg Astronomical Institute,
Universitetsky pr., 13, Moscow 119234 Russia
E-mail: ougolnikov@gmail.com

**Abstract**
Polarization measurements of the twilight background with Wide-Angle Polarization Camera (WAPC) are used to detect the depolarization effect caused by stratospheric aerosol near the altitude of 20 km. Based on a number of observations in central Russia in spring and summer 2016, we found the parameters of lognormal size distribution of aerosol particles. This confirmed the previously published results of the colorimetric method as applied to the same twilights. The mean particle radius (about 0.1 μm) and size distribution are also in agreement with the recent data of *in situ* and space-based remote sensing of stratospheric aerosol. Methods considered here provide two independent techniques of the stratospheric aerosol study based on the twilight sky analysis.

**1. Introduction**

The history of stratospheric aerosol observations covers more than 50 years, after the first balloon measurements provided by Junge et al. (1961). Aerosol concentration reaches the maximum near the altitude of 20 km (the Junge layer). The long-lasting series of balloon experiments started in 1970s (Hofmann and Rosen, 1980) showed that stratospheric aerosol particles are the droplets of sulfur acid solution (Rosen, 1971). Their number sufficiently increases after intensive volcanic eruptions due to the emission of $SO_2$ into the stratosphere. A combined analysis of balloon and lidar measurements in 1970s described in (Swissler et al., 1982) showed the effects of several volcanoes, of which the most noticeable ones were Fuego in 1974 and St. Helens in 1980. The eruption of El Chichon in 1982 significantly changed the aerosol characteristics for the following five years (Hofmann and Rosen, 1987). During these years, stratospheric aerosol was studied by balloon techniques (Hofmann and Rosen, 1982), lidars (McCormick et al., 1984), and satellites (SAGE II, Thomason et al., 1997). The most recent major eruption, which was also the most powerful one in 20th century, was Mt. Pinatubo in 1991, whose significant effect on stratospheric aerosol lasted for several years (McCormick et al., 1995; Bauman et al., 2003; Deshler et al., 2003; Jäger, 2005).

In the beginning of 21th century, the effects of several minor eruptions were recorded by space-based techniques (SCIAMACHY, Von Savigny et al., 2015) and lidar remote sensing (Burlakov et al., 2011, Ridley et al., 2014). During the recent years, stratospheric aerosol remained in its background conditions, and it was the longest volcanically-quiet period since the beginning of regular stratosphere observations. However, its properties seem to change over time. A positive trend of background aerosol was noticed in (Solomon et al., 2011). Earlier measurements by Hofmann and Rosen (1980) had shown an increase in the stratospheric aerosol amount compared to the data of Junge et al. (1961). However, this comparison could be affected by the low accuracy of Junge's results. The 30-year survey (Deshler et al., 2006) did not show any statistically significant trends of stratospheric aerosol. This question is important, since background aerosol can have partially anthropogenic origin, related to urban emissions of carbonyl sulfide (Crutzen, 1976) and sulfur dioxide (Brock et al., 1995).

One of basic characteristics of stratospheric aerosol related to the physical processes of its formation that could be found by the observations is the size distribution of particles. Its determination was the fundamental goal of the balloon measurements (Deshler et al., 2003) and space-based analysis (Bingen et al., 2004). It was shown that under the background conditions, the size distribution is well described by lognormal function:

$$dn(r) = \frac{N}{\sqrt{2\pi} \ln \sigma} \exp\left(-\frac{\ln^2(r/r_0)}{2\ln^2 \sigma}\right) d\ln r, \qquad (1)$$

where $N$ is the total number of particles; $r_0$ is the mean radius; and $\sigma>1.0$ is the distribution width. In volcanically-perturbed conditions, the distribution becomes bimodal with the second peak characterized by larger $r_0$ and related to particles of volcanic origin. The typical values of background lognormal distribution parameters found by Deshler et al. (2003) and at other measurements are $r_0 \sim 0.08$ μm, $\sigma \sim 1.6$, being practically independent from the time. However, the mean radius depends on the altitude, with a maximum near 20 km, in both background and post-volcanic conditions (Hofmann and Rosen, 1982).

At the present time, the stratospheric aerosol characteristics are being measured within the OSIRIS experiment held onboard the Odin satellite (Bourassa et al., 2012) and OMPS LP experiment held onboard the NPP satellite (DeLand, 2017) using the limb viewing geometry. The light-scattering properties at different wavelengths can be compared to obtain the size distribution of aerosol particles. Bourassa et al. (2008) found that the size distributions are in good agreement with the ones described above.

An analysis of light scattering on stratospheric aerosol for the size determination can be performed from the ground using the multi-wavelength lidar technique (Di Girolamo et al., 1995); in this case, backscattering (phase angle 180°) is considered. It can be also done during the twilight, when the stratosphere is illuminated by solar radiation while the troposphere is not. A twilight analysis can not have a good altitude resolution; however, it can involve a wide range of phase angles, increasing the accuracy of the size distribution retrieval.

Ugolnikov and Maslov (2017) used the data on the twilight sky color (or intensity ratio at different wavelengths) to find the aerosol scattering component of the twilight background and to study the particle properties. Despite the approximate (or even empirical) nature of the method, the size distribution parameters were found to be close to the ones determined by the space-based or balloon experiments. In this paper, we use an independent approach to the same observational data and find the aerosol characteristics by the polarization analysis, comparing them with the color data. Twilight sky polarization is sensitive to aerosol scattering, and this fact helped to detect the aerosol from the Tavurvur (Rabaul) volcano in 2006 (Ugolnikov and Maslov, 2009), using just one-wavelength measurements near the zenith. The color and polarization analyses are found to be effective for noctilucent cloud study (Ugolnikov et al., 2016, 2017). The background stratospheric aerosol analysis is more difficult due to less contribution to the total sky brightness; however, multi-wavelength measurements over a large part of the sky make it possible.

**2. Observations and polarization effects of aerosol**

Multi-color polarization measurements of the twilight sky background were conducted in Chepelevo (55.2°N, 37.5°E, 50 km southwards from Moscow) in March–July 2016. The observations were performed with a Wide-Angle Polarization Camera (Ugolnikov and Maslov, 2013ab). The field diameter was 140°; the points under consideration were at the zenith angles of up to 55°–60°; RGB-color CCD Sony QHYCCD-8 was used; the effective wavelengths were equal to 461, 540, and 624 nm for wide RGB channels, respectively. The R channel was corrected with an IR-blocking filter with the threshold at 680 nm. The exposure times varied from 3 ms to 30 s, depending on the twilight stage. Star images at different zenith angles on the night frames were analyzed to identify the camera position, flat field, and atmosphere transparency.

The procedure considered herein is quite similar to the one based on the same observations and described in (Ugolnikov and Maslov, 2017). However, we took polarization and its difference in R and B bands instead of the sky intensity and color. We use the polarization data in the solar vertical; the point position is characterized by the zenith angle $\zeta$, which is positive in the dusk area and negative in the opposite part of the sky. The solar zenith angle was denoted as $z$.

When describing the polarization properties of the sky background in the solar vertical, it is sufficient to find the normalized second Stokes component:

$$q(\zeta, z) = -\frac{I_2(\zeta, z)}{I(\zeta, z)} = -p(\zeta, z)\cos 2A(\zeta, z). \qquad (2)$$

Here, $I$ and $I_2$ are the first and second Stokes components, $p$ is the sky polarization degree, and $A$ is the polarization direction angle with respect to the solar vertical. During the light twilight the angle $A$ is close to 0° or 90° in solar vertical, so the quantity $q$ is usually equal to $\pm p$ and can be simply called the polarization. High above the horizon, the value of $q$ is positive due to the properties of Rayleigh scattering. It decreases and can be even negative closer to the horizon both in the dusk and opposite parts of the sky; the main reason is multiple scattering contribution (Ugolnikov, 1999). A change of the ratio of single and multiple scattering explains the general polarization properties of the twilight background, including the decrease in polarization at the solar zenith angle above 95–96° observed for any sky positions and wavelengths (Ugolnikov and Maslov, 2002, 2007, 2013ab). Aerosol scattering in a certain atmospheric layer causes additional depolarization effects visible during the special twilight stage, when effective scattering takes place in the same layer. This effect was revealed after the Rabaul volcano eruption in 2006 (Ugolnikov and Maslov, 2009). Due to weaker wavelength dependency compared to Rayleigh and multiple scattering, the aerosol effects are expected to be most noticeable in the red spectral band.

Figure 1 shows the dependencies of $Q$ on $z$ for different sky points ($\zeta$) and color bands (R, G, B). We can see that behavior of polarization is principally the same for all color bands in each certain sky point. Sky polarization is a little bit higher in the R band due to the lower contribution of multiple scattering (Ugolnikov et al., 2004). This difference remains almost constant until the solar zenith angle of about 97°, when single scattering vanishes fast against the background of multiple scattering.

However, there is another effect seen during the light twilight, at the solar zenith angle of about 93°: polarization depression in the R band that is not noticed in the B band. The effect is practically unseen in the dusk-opposite region of the sky, where the curves corresponding to different bands remain parallel. But it appears near the zenith and becomes stronger in the dusk area. It is normal for Mie scattering to be excessive at low phase angles and have less polarization compared to Rayleigh scattering. This depolarization takes place when the effective altitude of scattering is about 20 km, that corresponds to the Junge layer. During the same period, the brightness excess in the red band appears in the dusk segment. Thus, both effects can be interrelated with stratospheric aerosol. Here, we use the polarization effect to find the aerosol characteristics and compare the results of color and polarization procedures of the stratospheric aerosol study.

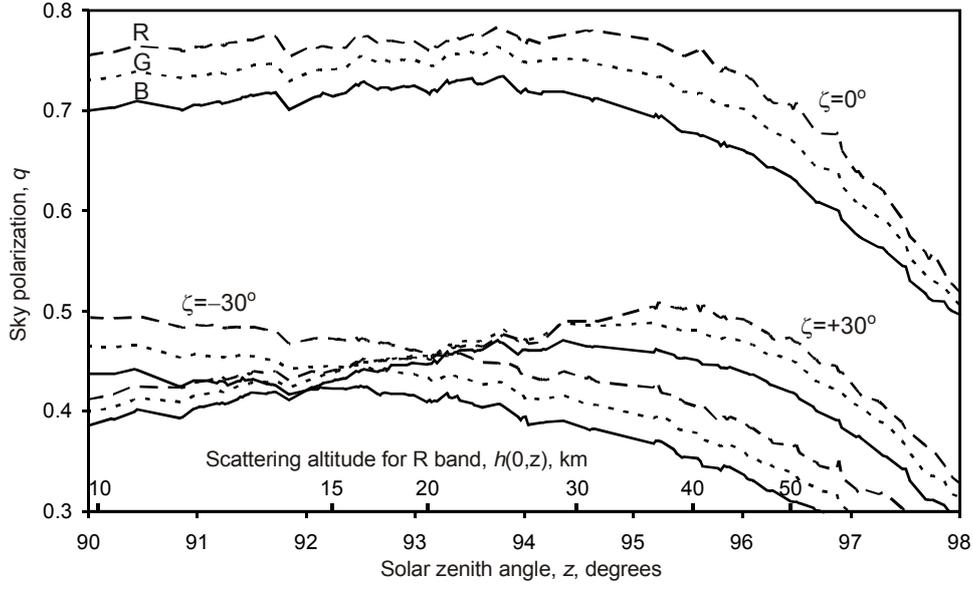

*Figure 1. Sky background polarization in different spectral bands and solar vertical points depending on solar zenith angle and effective scattering altitude, evening twilight of March, 27, 2016.*

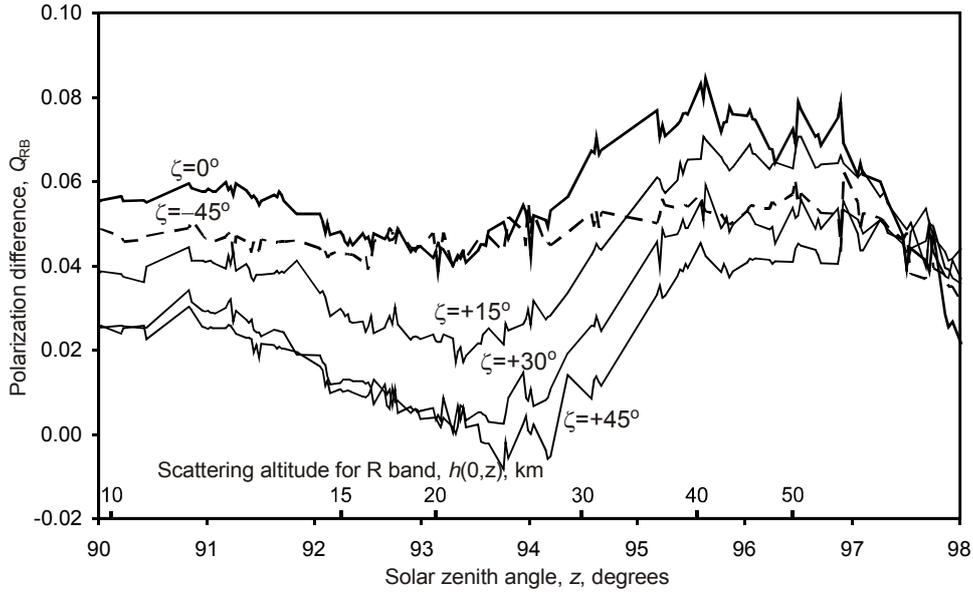

*Figure 2. Polarization difference in R and B bands for different solar vertical points, the same twilight as in Figure 1.*

**3. Study of aerosol scattering based on the polarization data**

The procedure of the analysis of the aerosol scattering contribution in the R band is quite similar to the procedure described in (Ugolnikov and Maslov, 2017), with polarization replacing the intensity logarithm. We introduce the value of difference of polarization in the R and B bands:

$$Q_{RB}(\zeta, z) = q_R(\zeta, z) - q_B(\zeta, z). \tag{3}$$

Figure 2 shows the dependence of $Q_{RB}$ on solar zenith angle $z$ for different points of the solar vertical. The deep twilight interval of $z$ between 95.5° and 97.5° is characterized by the almost constant polarization difference. It is the period when single scattering takes place in the upper stratosphere and lower mesosphere with a negligibly small amount of aerosol. The polarization difference remains practically the same during the light twilight in the dusk-opposite region (the curve for $\zeta=-45°$ is shown in Figure 2) but starts falling near the zenith and, especially, in the dusk sky area. In the same time and sky part, background color turns redder. Both effects are related to stratospheric aerosol below 40 km, scattering the solar radiation basically in the forward direction.

Following (Ugolnikov and Maslov, 2017), we take moment $z_0=96°$ as the reference and assume that in case of aerosol-free atmosphere, value $Q_{RB}$ would change slowly and uniformly along the solar vertical during the twilight. Analyzing the change of polarization difference, $Q_{RB}(\zeta, z) - Q_{RB}(\zeta, z_0)$, we can see its dependency on $\zeta$ for the certain solar zenith angle $z$ in Figure 3. The depolarization effect increasing in the dusk segment is worthy of note. Following the color method again and based on Figure 1, we assume that the stratospheric aerosol contribution in the B band is sufficiently less than in the R band. This assumption is based not only on the lower ratio of aerosol and Rayleigh scattering in the B band, but also on the higher effective altitude of single scattering in the B band at the same moment. It causes the underestimation of the aerosol contribution, but it is effective for size distribution study. We denote the brightness and polarization of the aerosol-free background in the R band as $I_{0R}$ and $q_{0R}$; the same values for the aerosol component as $I_{AR}$ and $q_{AR}$. For the total background, we have:

$$I_R(\zeta, z) = I_{0R}(\zeta, z) + I_{AR}(\zeta, z);$$
$$I_R q_R(\zeta, z) = I_{0R} q_{0R}(\zeta, z) + I_{AR} q_{AR}(\zeta, z). \quad (4)$$

Aerosol scattering decreases the total sky polarization by the following value:

$$q_R(\zeta, z) - q_{0R}(\zeta, z) = a(\zeta, z) (q_{AR}(\zeta, z) - q_{0R}(\zeta, z));$$
$$a(\zeta, z) = I_{AR}(\zeta, z) / I_R(\zeta, z). \quad (5)$$

Here, $a$ is the contribution of aerosol scattering in the total sky background in R band. Following our assumption, the change in the aerosol-free sky polarization difference in the R and B bands is slow by $z$ and uniform along the solar vertical:

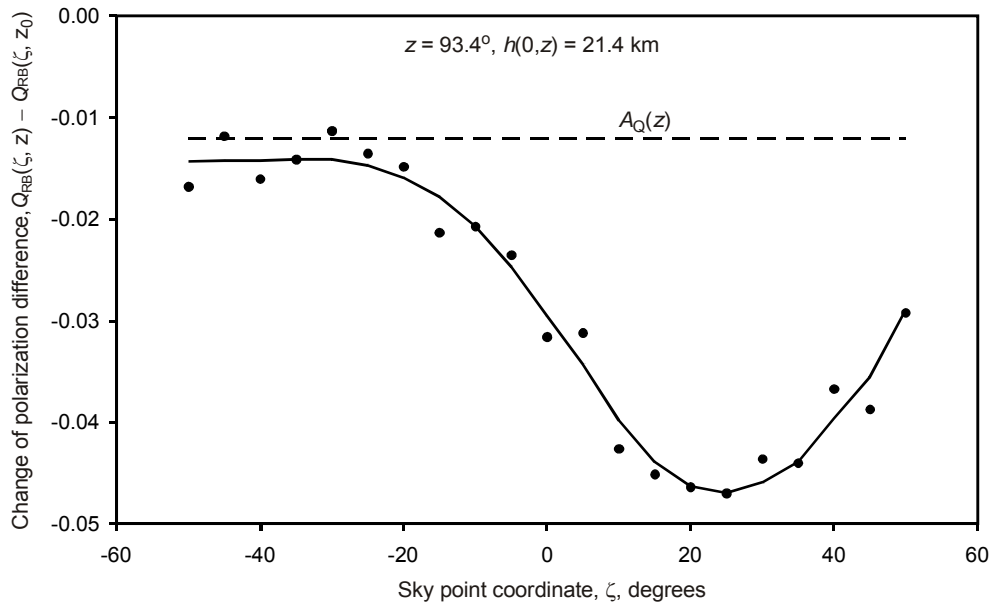

Figure 3. Example of the procedure of analysis of aerosol scattering basing on the polarization data.

$$(q_{0R}(\zeta, z) - q_B(\zeta, z)) - (q_{0R}(\zeta, z_0) - q_B(\zeta, z_0)) = A_Q(z). \quad (6)$$

For the deep twilight ($z=z_0$) $q_{0R} = q_R$. Substituting (6) into (5), we have:

$$q_R(\zeta, z) = a(\zeta, z)(q_{AR}(\zeta, z) - q_{0R}(\zeta, z)) + q_B(\zeta, z) + A_Q(z) + (q_R(\zeta, z_0) - q_B(\zeta, z_0)). \quad (7)$$

Using the definition (3), we rewrite it as follows:

$$Q_{RB}(\zeta, z) = Q_{RB}(\zeta, z_0) + A_Q(z) + a(\zeta, z)(q_{AR}(\zeta, z) - q_{0R}(\zeta, z)). \quad (8)$$

Here, $Q_{RB}$ values are measured during the observations. The contribution of aerosol scattering in the total background is equal to:

$$a(\zeta, z) = \frac{I_{AR}(\zeta, z)}{I_R(\zeta, z)} = \frac{F(r_0, \sigma, z-\zeta) P(h(\zeta, z))}{\cos\zeta \cdot I_R(\zeta, z)} e^{-\tau} e^{\frac{-E_R}{\cos\zeta}}. \quad (9)$$

Here, $F$ is the first component of the Mie scattering matrix of sulfate particles with lognormal distribution with mean radius $r_0$ and distribution width $\sigma$; the scattering angle is equal to ($z-\zeta$); refraction can be disregarded here. The refractive index $m$ of sulfur acid droplets for the stratosphere is 1.44 (Russell and Hamill, 1984). Ugolnikov and Maslov (2017) used the value 1.43 for the color analysis, which did not change the results; the difference of the mean radius is less than 0.0005 μm. However, the color results used herein for comparison were recalculated for the refractive index 1.44. Parameters $m$, $r$, and $\sigma$ also define value $q_{AR}$. $P$ is the vertical profile of aerosol, $h$ is the effective scattering altitude, $\tau$ is the optical depth of the light trajectory before scattering, $E_R$ is the vertical optical depth of the atmosphere in the R band determined experimentally.

We take the $h$ value as corresponding to the solar emission path to the scattering point with optical depth $\tau=1$. The constant term $e^{-\tau}$ can be included to the value of $P$ determined in arbitrary units. We calculate altitude $h$ using the atmospheric model with real temperature and ozone vertical profiles for each observation date by EOS Aura/MLS data (EOS Team, 2011ab). These values for zenith are denoted along the $x$-axis in Figures 1 and 2. This altitude depends on the sky point, decreasing to the dusk area. For each corresponding interval, dependence $P(h)$ is assumed to be exponential:

$$P(h(\zeta, z)) = P_0(h(0, z)) e^{-K(h(\zeta, z) - h(0, z))}. \quad (10)$$

Finally, we have:

$$Q_{RB}(\zeta, z) - Q_{RB}(\zeta, z_0) = A_Q(z) +$$
$$+ \frac{(q_{AR}(r, \sigma, z-\zeta) - q_{0R}(\zeta, z)) F(r_0, \sigma, z-\zeta) P_0(h(0, z)) e^{-K(h(\zeta, z) - h(0, z))}}{\cos\zeta \cdot I_R(\zeta, z)} e^{\frac{-E_R}{\cos\zeta}}. \quad (11)$$

Having a number of measurements for fixed $z$ and different $\zeta$ (the step by $\zeta$ is 5°), we consider it as an equation system with unknown parameters $A_Q$, $r_0$, $\sigma$, and $P_0$. Initially, we do not know the value of the aerosol-free background polarization in the R band, $q_{0R}$. For the first approximation, we assume it to be equal to:

$$q^{(0)}{}_{0R}(\zeta, z) = q_B(\zeta, z) + (q_R(\zeta, z_0) - q_B(\zeta, z_0)). \quad (12)$$

This would be true in the case $A_Q(z)=0$ in equation (7). It is close to reality, as is seen in Figure 3. We also do not know the gradient of the vertical aerosol distribution, $K$, initially assuming it to be the same as for the Rayleigh scattering. Then, we can solve the system by the least squares method. A graphical example of this procedure is shown in Figure 3. When it is done, we find value $A_Q(z)$ and find the aerosol-free polarization value in the R band:

$$q_{0R}(\zeta, z) = q_B(\zeta, z) + (q_R(\zeta, z_0) - q_B(\zeta, z_0)) + A_Q(z). \quad (13)$$

Solving the system for different solar zenith angles and corresponding scattering altitudes, we find the gradient of the aerosol altitude profile:

$$K(h) = -\frac{dP_0(h(0,z))}{P_0(h(0,z))dh}. \quad (14)$$

Using the found values of $q_{0R}$ and $K$, we solve the system (11) again, making the next step of the iteration procedure. After several steps, we find the size distribution parameters, $r_0$ and $\sigma$. The process is very similar to the color analysis in (Ugolnikov and Maslov, 2017). However, in this case, there is no need to assume the aerosol contribution value $a$ to be low and take into account the color effect of the Chappuis bands of atmospheric ozone, as was done in the mentioned study.

## 4. Results and discussion

In any case of optical remote sensing, best-fit parameters ($r_0$, $\sigma$) can form a narrow region or a line on diagram ($r_0-\sigma$); wider distributions (more $\sigma$) correspond to lower mean radii $r_0$. A sample of distributions fitting the data of the twilight on March 27, 2016, at the altitude of 21.4 km are shown in Figure 4. They have similar right-side slopes, that is logical since the largest particles make the fundamental contribution in light scattering. Dependences of polarization of scattered light $q_{AR}$ on the phase angle are shown in Figure 5, they are very close to each other.

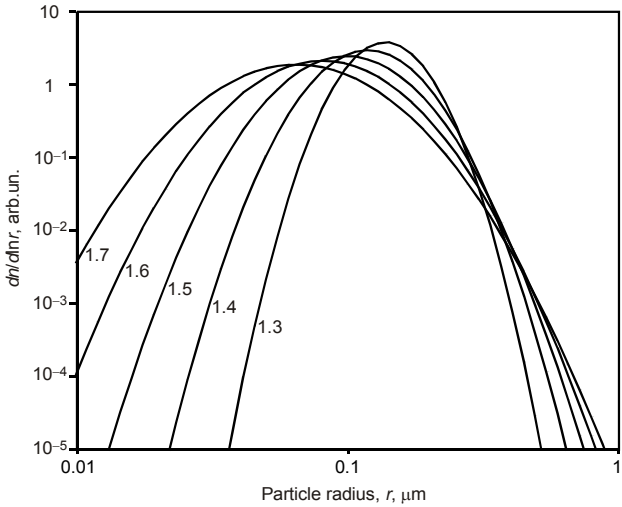
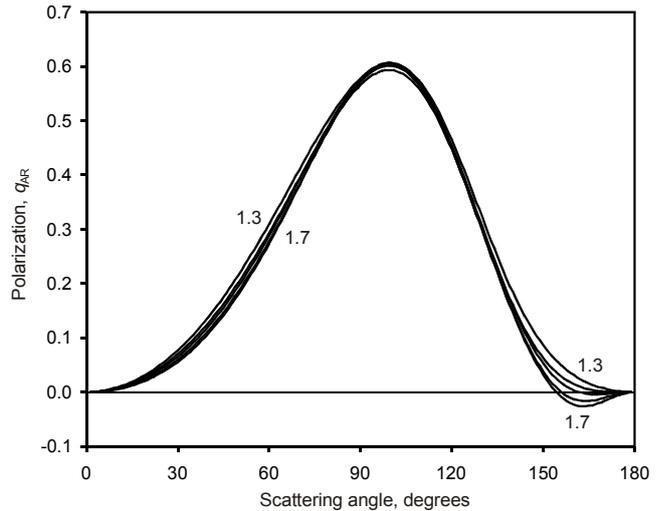

*Figure 4. Lognormal particle size distributions having close polarization properties of light scattering. Values of $\sigma$ are shown.*

*Figure 5. Angular dependencies of polarization of scattered light for particle size distributions in Figure 4, wavelength 624 nm. Values of $\sigma$ are shown.*

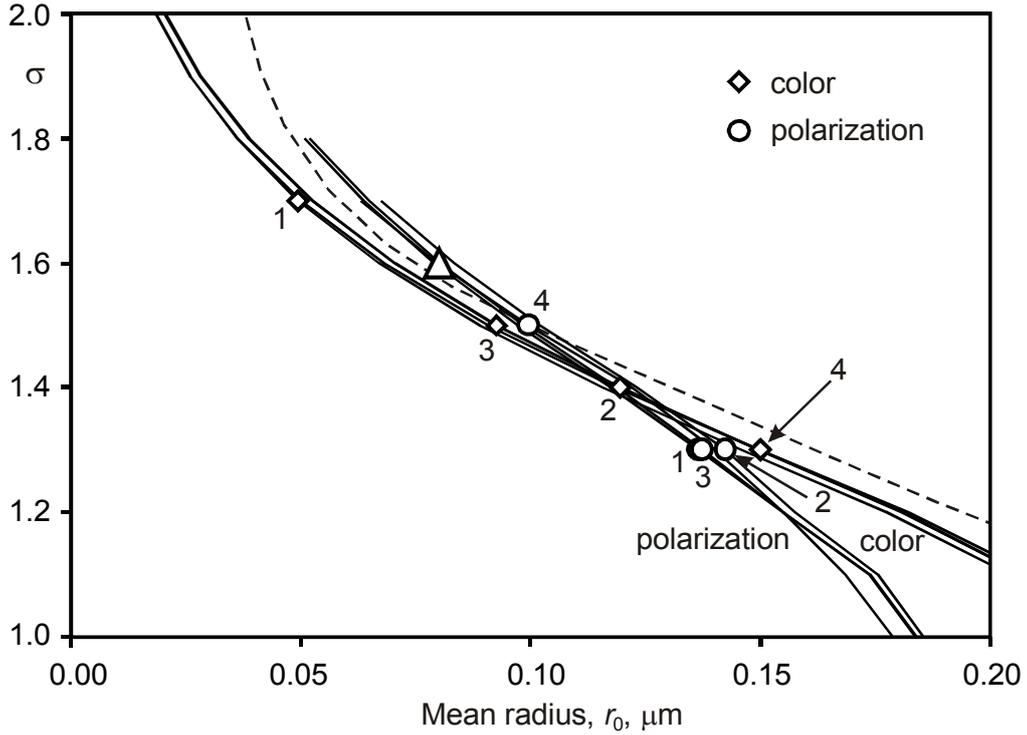

*Figure 6. Retrieved parameters of lognormal size distribution of aerosol using color and polarization methods, twilight of March, 27, 2016, altitudes 15.3 km (1), 18.6 km (2), 21.4 km (3), 24.3 km (4). Triangle symbol corresponds to typical distribution from (Deshler et al., 2003), dashed line – (Bourassa et al., 2008), 18 km.*

However, the absolute best-fit solution ($r_0$, $\sigma$) can be also found. It is shown in Figure 6: the results of the color and polarization analysis of four twilight moments (or effective scattering altitudes in the zenith) of the same twilight are plotted. We can see that lines ($r_0$–$\sigma$) are not the same for the color and polarization analysis, that helps to localize the area of exact solutions. Best-fit pairs ($r_0$, $\sigma$) are found to lie near in this area, especially for the polarization procedure. Most observation results relate to narrower size distributions ($\sigma$=1.4) than those found in (Deshler et al., 2003) and plotted in the same figure, but they are also placed almost on the same lines ($r_0$–$\sigma$). Satellite measurements (Bourassa et al., 2008) also result in a close line ($r_0$–$\sigma$).

Figure 7 shows the vertical profiles of the mean particle size with two assumed values of the size distribution width: 1.4 found herein (twilight on March 27) and 1.6 determined in (Deshler et al., 2003) and assumed in (Bourassa et al., 2008). On the plot, the results of the color and polarization analysis are provided, as well as the mean size profile found by (Bourassa et al., 2008). The results are in good agreement; the most remarkable feature is the size maximum at the altitude of about 22 km. However, it is blurred in the twilight data due to the restricted altitude resolution of the twilight analysis. At any twilight moment, the effective single scattering takes place across a wide range of altitudes.

The altitude profiles of values $F \cdot P$ in two solar vertical points (arbitrary units) defined by the color and polarization analysis for the same twilight are plotted in Figure 8. The data is compared with the OMPS and OSIRIS satellite profiles on aerosol extinction (525 nm) for the same date and location. We can see that the twilight profiles are blurred again for the same reasons; the altitude dependence is close to exponential. However, the mean altitude gradient of aerosol is the same. Multiple scattering, extinction and Chappuis absorption makes difficult to find the total amount of aerosol in absolute units based on the twilight analysis.

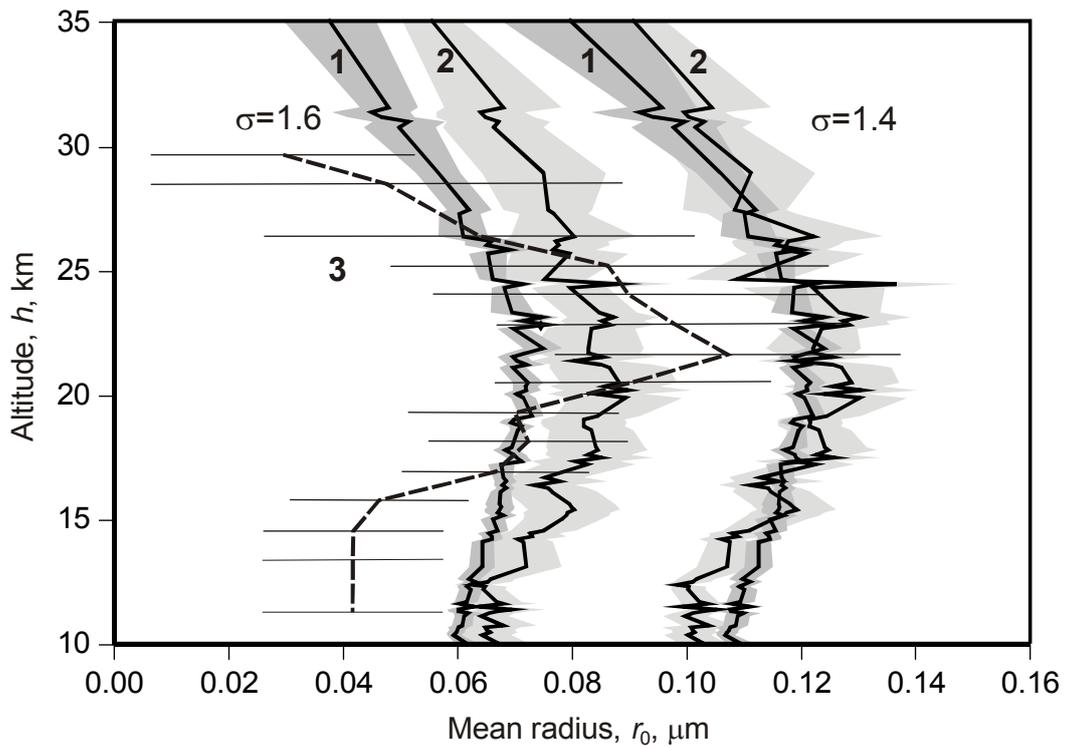

*Figure 7. Vertical profiles of mean radius with assumed distribution width σ obtained by color (1) and polarization (2) analysis compared with results of Bourassa et al. (2008), σ=1.6 (3).*

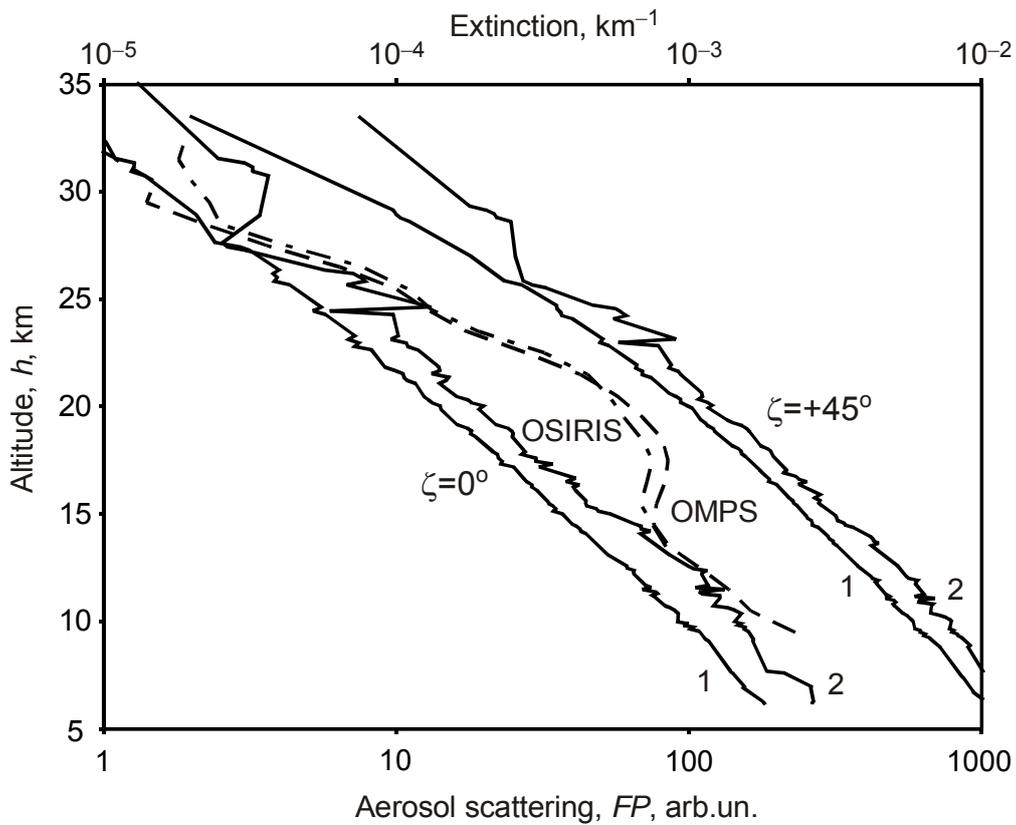

*Figure 8. Vertical profile of aerosol scattering in R band (1 – color analysis, 2 – polarization analysis) compared with OSIRIS and OMPS extinction profiles for March, 27, 2016, 525 nm.*

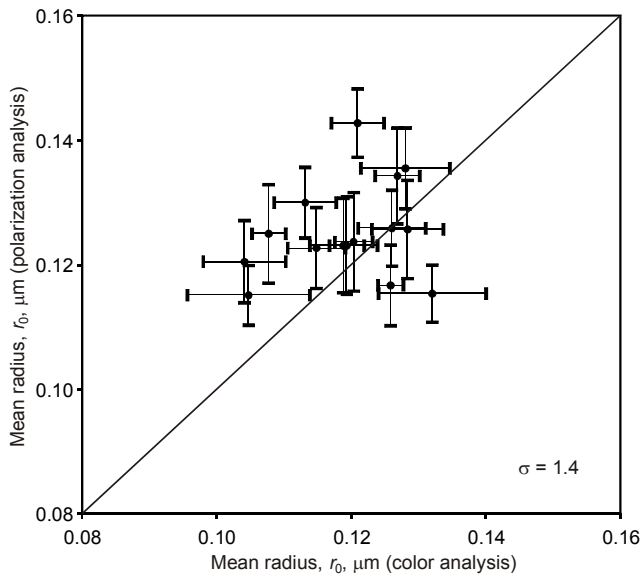 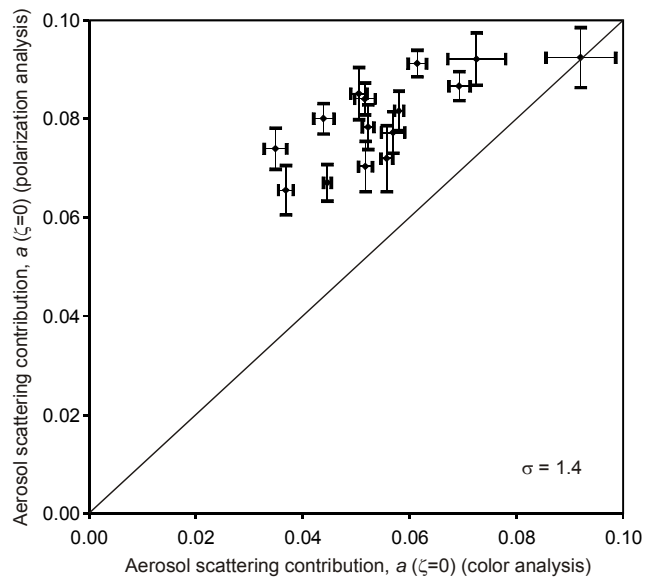

*Figure 9. Comparison of mean radius values for 20 km obtained by two methods, spring and summer twilights of 2016.*

*Figure 10. Comparison of aerosol scattering contribution in the twilight background (zenith, effective scattering altitude 20 km) obtained by two methods, spring and summer twilights of 2016.*

Finally, Figures 9, 10 present comparative diagrams of mean particle size $r_0$ ($\sigma=1.4$ is assumed) and aerosol contribution in the total sky background, $a$, in the zenith for the effective scattering altitude of 20 km, obtained by the color and polarization analysis of the same twilights in spring and summer 2016. We can see that the polarization analysis gives slightly higher values of aerosol contribution in the twilight background with less scattered results. However, this difference vanishes if we assume a wider size distribution. The mean particle radius values are quite similar.

## 5. Conclusion

The basic aim of this paper is to compare the stratospheric aerosol characteristics obtained during the same twilights by two independent techniques: an analysis of the sky color and polarization changes. Agreement of these results between each other and existing satellite data shows that the twilight sky analysis can be an effective way to hold a continuous survey of stratospheric aerosol by all-sky cameras, which seems to be a cost-effective technique. Good accuracy and a significant observational effect of background aerosol show that such measurements can be even more effective in the case of volcanically disturbed atmosphere. The twilight analysis does not have a good altitude resolution; however, it can provide the data on the size distribution and changes in the total amount of aerosol in the stratosphere.

The quality of the polarization results seems to be slightly better, but such measurements require a more complicated optical design of the camera. A color analysis can be performed with a simple all-sky ("fish-eye") lens with a color CCD. A large number of such cameras is now installed in northern latitudes for regular aurora survey; they can also be used for the microphysical study of noctilucent clouds (Ugolnikov et al., 2017). The stratospheric aerosol study can be another application of such devices. It is especially interesting due to changes of the stratospheric aerosol properties in the north polar vortex (McCormick et al., 1983) and possible appearance of polar stratospheric clouds. Long arrays of data can provide additional information on possible trends of stratospheric characteristics and influence of anthropogenic factors on this process.

**Acknowledgments**

Authors are thankful to Andrey M. Tatarnikov (Sternberg Astronomical Institute, Moscow State University, Russia) for his help in observations, Fedor V. Stytsenko (Space Research Institute, Russian Academy of Sciences) for his help during the data analysis, Landon Rieger and Adam Bourassa (University of Saskatchewan, Canada) for providing the OMPS and OSIRIS satellite data on stratospheric aerosol. The work is supported by Russian Foundation for Basic Research, grant No. 16-05-00170-a.